\documentclass[11pt,twoside]{article}

\usepackage{asp2004}
\usepackage{epsf}
\usepackage{lscape}
\usepackage[dvips]{graphicx}

\begin{document}
\title{LIRIS multi-slit \boldmath H$\alpha$ spectroscopy of a \boldmath $z\sim1$ DEEP2 \\
sample of star-forming galaxies}
\author{Nayra Rodr\'iguez-Eugenio\altaffilmark{1}, Kai G. Noeske\altaffilmark{2}, 
Jose Acosta-Pulido\altaffilmark{1}, Rafael Barrena\altaffilmark{1}, 
Francisco Prada\altaffilmark{3}, Arturo Manchado\altaffilmark{1, 4} and EGS Teams}
\altaffiltext{1}{Instituto de Astrof\'isica de Canarias, V\'ia L\'actea, E-38205 La Laguna, Tenerife, Spain.}
\altaffiltext{2}{Lick Observatory, University of California, 1156 High Street, Santa Cruz, CA 95064, USA.}
\altaffiltext{3}{Instituto de Astrof\'isica de Andaluc\'ia (CSIC), E-18008 Granada, Spain.}
\altaffiltext{4}{CSIC, Spain.}

\begin{abstract}
We present preliminary results of H$\alpha$ near-infrared 
spectroscopy of six galaxies with redshifts  
$z\sim1$ drawn from the DEEP2 Galaxy Redshift Survey. 
The spectra have been taken with the multi-slit mode of LIRIS 
(Long-slit Intermediate Resolution Infrared Spectrograph) 
installed at the 4.2-m William Herschel Telescope. 
This is a pilot study for a larger program to obtain 
H$\alpha$ luminosities of about 50 star-forming galaxies at $z\sim1$, 
with the aim of deriving the corresponding star formation rates 
(SFRs) from H$\alpha$ as well as studying the relationship with 
other SFR indicators. 
The new galaxy sample will be selected from the Extended Groth Strip Survey, 
where galaxies will also have measures of stellar masses, reddening, 
far-IR data, and galaxy morphologies.
\end{abstract}

\keywords{techniques: spectroscopic - infrared: galaxies - galaxies: formation - 
galaxies: evolution - galaxies: high-redshift}

\section{Introduction}

Several diagnostic methods have been used to measure star formation rates 
(SFRs) in galaxies, of which some of the most extensive are the rest-frame 
ultraviolet (UV) continuum (1500-2800 \AA), H$\alpha$ recombination line,
[O II]$\lambda3727$ forbidden line, and the far-IR luminosity.
These various SFR indicators are differently affected by dust extinction
and by metallicity effects, and they are related to the emission of different 
stellar populations, leading to discrepancies between the corresponding SFRs
measurements even after extinction corrections.
Therefore, in order to make a reliable study of the global star 
formation history of the Universe, it is necessary to use
the same SFR tracer from low to high redshifts, or to have 
good calibrations between different indicators.

The H$\alpha$ luminosity is an excellent direct tracer of the ``instantaneous'' SFR,
it is immune to metallicity effects, and 
dust obscuration affecting H$\alpha$ can be mitigated by reddening correction.
H$\alpha$ has been a common SFR tracer in low-$z$ surveys, 
but at moderate redshifts there are only small statistical galaxy samples 
with H$\alpha$ measurements (e.g.~Glazebrook et al.~1999;  
Tresse et al.~2002; Erb et al.~2003; Doherty et al.~2004; Shapley et al.~2005).
At $z>0.6$ H$\alpha$ is redshifted into the near-IR (NIR), 
where the high sky background entails the need of 
large total exposure times to obtain good spectral data of faint objects.
Only with the recent development of multi-object NIR spectrographs
we are able to obtain H$\alpha$ spectra of a significant number of 
high-$z$ galaxies in realistic observing periods.

The goals of the LIRIS-EGS project are to probe the star formation history
of the Universe at $z\sim1$ using H$\alpha$ luminosities as SFR indicators, 
to obtain self-consistent redshift-one SFR calibrations between H$\alpha$ 
and other star formation tracers, and to complete this study with galaxy 
stellar masses, morphologies, metallicities and reddening 
estimations.\footnote{This research is partly funded by the 
Spanish Min.~de Educacion y Ciencia (AYA2004-03136).}
This project is possible thanks to the multi-slit mode of the NIR 
spectrograph LIRIS (Long-slit Intermediate Resolution Infrared Spectrograph),  
which allows simultaneous observations of about 15 galaxies, and thanks to
the collaboration with the EGS teams.
In this work we present the results of a pilot study where we wanted
to probe the multi-slit capabilities of LIRIS at the 4.2-m William Herschel 
Telescope (WHT) to study the SFR properties of galaxies at $z\sim1$ from 
H$\alpha$ spectroscopy. 
The cosmology $\Omega_M = 0.3$, $\Omega_\Lambda = 0.7$, $H_0 = 70$ km s$^{-1}$ Mpc$^{-1}$
is assumed throughout.

\section{Target Selection, Observations and Data Reduction}
The sample of galaxies was drawn from the DEEP2 Redshift Survey \citep{dav03},
and they are objects for which we expected to find H$\alpha$ emission in the J-band,
inferred from the presence of [O II] in emission in the DEEP2 Keck optical spectra.
We pre-selected galaxies with redshifts in the range $0.8 \la z \la 1.3$ 
such that H$\alpha$ will appear in regions of good atmospheric transmission,
with good quality of the redshift estimations (90\% reliable), and with
[O II]$\lambda$3727 rest-frame equivalent widths larger than 20 \AA. 
We did not apply a selection criterion based on OH sky lines to study how
the presence of these lines near H$\alpha$ affects its detection.
The final selection consists of six DEEP2 $z\sim1$ galaxy targets 
within the $4.2\times4.2$ arcmin$^2$ LIRIS field of view 
centred at RA=$16^h 47^m 39\fs6$, Dec=34\deg 43\arcmin 48\arcsec (J2000).
The LIRIS multi-slit mask designed to observe these objects also includes holes for
two acquisition stars. 

The spectroscopic observations were carried out on 2005 June 22 with 
the multi-object mode of LIRIS at the 4.2-m WHT.
LIRIS is a NIR (0.9-2.4 \micron) intermediate resolution
spectrograph, with a Hawaii-II detector cooled at $\sim$
70 K, and a pixel scale of 0.25 arcsec \citep{man00}.
We used a zJ grism and a slit width of 0.9 arcsec, 
yielding a dispersion of 6 \AA \hspace{0.5mm} per pixel, and a resolution
$\lambda / \Delta \lambda \sim 600$. 
During the observations the telescope was nodded between three positions 
along the slit, with a separation of 4 arcsec between
contiguous positions. The total exposure time was 3.5 hours divided 
into 7 series of three 600 sec exposures, ensuring that we are limited 
by photon noise.

\begin{figure}[!t]
\begin{center}
\includegraphics[width=10.2cm]{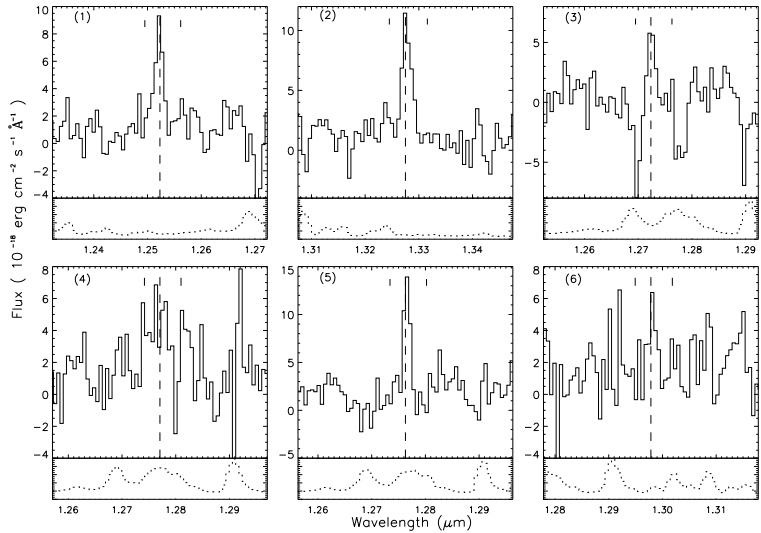}
\caption{\label{fig:multispec}\footnotesize{Observed-frame spectra of our six LIRIS/DEEP2 galaxies. 
The expected position for H$\alpha$ given the optical spectroscopic redshift
is marked with a vertical dashed line, and the vertical bars to either side
show the predicted positions of the [N II]$\lambda$6548 and [N II]$\lambda$6583 lines. 
Plotted below each galaxy spectrum is the subtracted night sky spectrum in arbitrary 
flux units. The OH sky lines are not resolved in these spectra.}}
\end{center}
\end{figure}

Our spectroscopic reduction includes a first order
``extra-sky subtraction'' process, 
and wavelength and flux calibration.  
No dark or flat-field corrections were performed. In the first case the 
reason is the low temperature of the LIRIS detector, while the second decision
was made on the basis of the noisier results found after this correction, 
due to the absence of continuum in the galaxy spectra.
The reduced emission line spectra are shown in \figurename{} \ref{fig:multispec}.
We measured the H$\alpha$ integrated fluxes, associated errors, and FWHMs, 
by interactively fitting Gaussian functions to the background-subtracted 
emission-line profiles using the package \emph{fitlines} under IRAF
\citep{aco00}.

\section{Preliminary Results and Discussion}
We find four clear H$\alpha$ detections with S/N $>$ 4
(galaxies 1, 2, 3 and 5), and another 3-$\sigma$ detection 
(galaxy 6), out of the six observed galaxies.
Only the two galaxies with H$\alpha$ emission away from sky lines have FWHMs greater 
than or similar to the instrumental resolution. This result highlights 
the importance of a target selection criterion based on the distance
between H$\alpha$ and intense OH skylines, which has been included in the selection of
the new LIRIS-EGS sample.

\begin{figure}[!t]
\begin{center}
\includegraphics[width=8.7cm]{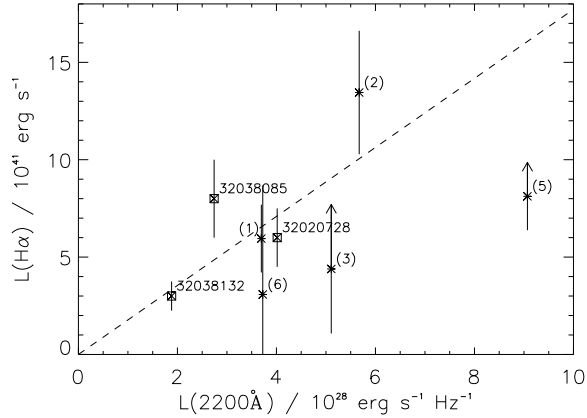}
\caption{\label{fig:LUVHa}\footnotesize{Comparison of the H$\alpha$ luminosity and the UV 2200\AA \hspace{0.1mm}
flux for the individual galaxies. The dashed line represents the relation 
between $L_\nu$ and $L_{H\alpha}$ to yield the same SFR using the Kennicutt
conversions. Starred symbols correspond to our galaxies and 
squares to the $z\sim1$ galaxies from \protect\citet{sha05}.
Arrows in galaxies (3) and (5) show that these luminosities are lower limits due to subtraction
of near intense sky lines.}}
\end{center}
\end{figure}

We obtained the H$\alpha$ luminosities, $L_{H\alpha}$, of the galaxies and derived
SFRs using the \citet{ken98} conversion (2).
As our galaxies have redshifts near unity, we directly derived UV luminosity densities, 
$L_\nu$(2200 \AA), from the DEEP2 B magnitudes (AB) as a good first approximation.
We also derived UV SFRs using the \citet{ken98} conversion (1).
These magnitudes are summarized in \tablename{} \ref{tab:tbl_1} together with 
the DEEP2 identification numbers and redshifts ($z_{opt}$) of the galaxies, 
and redshifts derived from LIRIS H$\alpha$ detections ($z_{H\alpha}$).
In \figurename{} \ref{fig:LUVHa} we plot the H$\alpha$ versus UV luminosities of
our galaxies and of three $z\sim1$ DEEP2 galaxies from \citet{sha05}.

Our non-extinction corrected SFRs inferred from H$\alpha$ are smaller than the 
corresponding UV SFRs in all cases except one, and the same result is found 
for the \citeauthor{sha05} galaxies. Other studies of the UV - H$\alpha$
SFR relation (e.g.~Glazebrook et al.~1999; Doherty et al.~2004) 
had found opposite results, with understimations of the UV SFRs by 
a factor of 2 or 3 from not extinction-corrected data. One of the reasons for
this discrepancy is that four of the H$\alpha$ lines in our spectra are affected by 
the subtraction of very intense OH sky lines, which could reduce the H$\alpha$ fluxes.
Also, the galaxies appear to be extended in the I-band DEEP2 images, with a mean FWHM 
of 1.4 arcsec, so we need to apply aperture corrections to make up for our 0.9 arcsec 
wide slits. 

Even though final results are not achieved, with this pilot study we have demonstrated 
that LIRIS is highly suitable for obtaining H$\alpha$ spectroscopy of a significant
number of $z\sim1$ star-forming galaxies. With the upcoming LIRIS-EGS sample we will 
have a comprehensive dataset that includes not only
 H$\alpha$ and UV continuum information, but also [O II] and far-IR luminosities, 
stellar masses, reddening estimations, galaxy morphologies and metallicities.

\begin{table}[!hb]
\caption{\label{tab:tbl_1}\footnotesize{Observed galaxy properties derived from DEEP2 and LIRIS data}}
\smallskip
\begin{center}
{\scriptsize
\begin{tabular}{cccccccc}
\tableline
\noalign{\smallskip}
id &DEEP2 id & $z_{opt}$ & $z_{H\alpha}$ & $L_\nu$ (2200 \AA) & SFR$_{UV}$  & $L_{H\alpha}$ & SFR$_{H\alpha}$ \\ 
 & & & &  ($10^{28}$ erg s$^{-1}$Hz$^{-1}$)  & (M$_{\sun}$ yr$^{-1})$  & ($10^{41}$ erg s$^{-1}$)  & (M$_{\sun}$ yr$^{-1})$\\
\noalign{\smallskip}
\tableline
\noalign{\smallskip}
1 & 21002068 &  0.90820 &  0.9079 &  3.69 &  5.2 &  5.95 $\pm$ 1.74 &	4.7 $\pm$ 1.4 \\

2 & 21006790 &  1.02276 &  1.0231 &  5.67 &  7.9 & 13.45 $\pm$ 3.17 & 10.6 $\pm$ 2.5 \\

3 & 21002050 &  0.93875 &  0.9388 &  5.11 &  7.1 &  4.39 $\pm$ 3.31 &	3.5 $\pm$ 2.6 \\

4 & 21006616 &  0.94590 &    -    &  3.93 &  5.5 &     -            &      -         \\

5 & 21006922 &  0.94472 &  0.9451 &  9.07 & 12.7 &  8.12 $\pm$ 1.74 &	6.4 $\pm$ 1.4 \\

6 & 21006954 &  0.97758 &  0.9780 &  3.72 &  5.2 &  3.08 $\pm$ 5.60 &	2.4 $\pm$ 4.4 \\
\noalign{\smallskip}
\tableline
\end{tabular}
}
\end{center}
\end{table}

\end{document}